\documentclass[groupedaddress,showpacs,aps,preprint,prb,floatfix]{revtex4-2}
\usepackage{graphicx}
\usepackage{textcomp}
\usepackage{amsmath}

\begin{document}

\title{Charge Transfer Dynamics in an Electron-Hole Bilayer Device: Capacitance Oscillations and Hysteretic Behavior}

\author{M. L. Davis}
\affiliation{Solid State Physics Laboratory, ETH Z\"urich, CH-8093 Z\"urich, Switzerland}
\affiliation{Quantum Center, ETH Z\"urich, CH-8093 Z\"urich, Switzerland}
\author{S. Parolo}
\affiliation{Solid State Physics Laboratory, ETH Z\"urich, CH-8093 Z\"urich, Switzerland}
\author{S. Agostini}
\affiliation{Solid State Physics Laboratory, ETH Z\"urich, CH-8093 Z\"urich, Switzerland}
\author{C. Reichl}
\affiliation{Solid State Physics Laboratory, ETH Z\"urich, CH-8093 Z\"urich, Switzerland}
\affiliation{Quantum Center, ETH Z\"urich, CH-8093 Z\"urich, Switzerland}
\author{W. Dietsche}
\affiliation{Solid State Physics Laboratory, ETH Z\"urich, CH-8093 Z\"urich, Switzerland}
\affiliation{Max-Planck-Institut f\"ur Festk\"orperforschung, D-70569 Stuttgart, Germany}
\author{W. Wegscheider}
\affiliation{Solid State Physics Laboratory, ETH Z\"urich, CH-8093 Z\"urich, Switzerland}
\affiliation{Quantum Center, ETH Z\"urich, CH-8093 Z\"urich, Switzerland}

\begin{abstract}
The capacitance and differential conductance of MBE-grown AlGaAs/GaAs p-i-n diodes are investigated. In these diodes, the p-doped layer, an adjacent intrinsic spacer, and a central barrier are made of AlGaAs. Capacitance oscillations and hysteretic behavior are observed and understood to be consequences of the AlGaAs spacer properties. These findings have significant implications for the design of heterostructures aimed at achieving electrically contacted, closely spaced electron and hole layers.
\end{abstract}

\maketitle


\section{Introduction}

Electron-hole bilayers (EHBs) are theoretically predicted to host various excitonic phases at different temperatures, densities, and interlayer spacings \cite{Snoke2011,Fogler2014,DasSarma2024}. Consequently, they have attracted significant research interest for decades \cite{butov2002,Ma2021,shaibley2025}. GaAs/AlGaAs heterostructures are one of the principal material systems in this investigation \cite{seamons2009,cumis2017,dubin2021,choksy2023}, for which very high material qualities are achievable, and established electrical contacting methods exist. In this material system, EHBs can be generated electrically in order to study their transport properties. To access excitonic regimes, a key challenge is to grow heterostructures in which the interlayer spacing is sufficiently small. This is difficult due to the tendency for interlayer current to increase dramatically with decreasing interlayer spacing. In addition, high-quality contacts to the individual layers must be established. It therefore becomes essential to understand the interlayer transport characteristics of these devices, with the aim of having closely spaced EHBs with negligible charge transfer between the layers, but simultaneously negligible resistance from the charge layers to the device contacts.

\begin{figure}[!t]
	\centering
	\includegraphics[width=0.5\columnwidth]{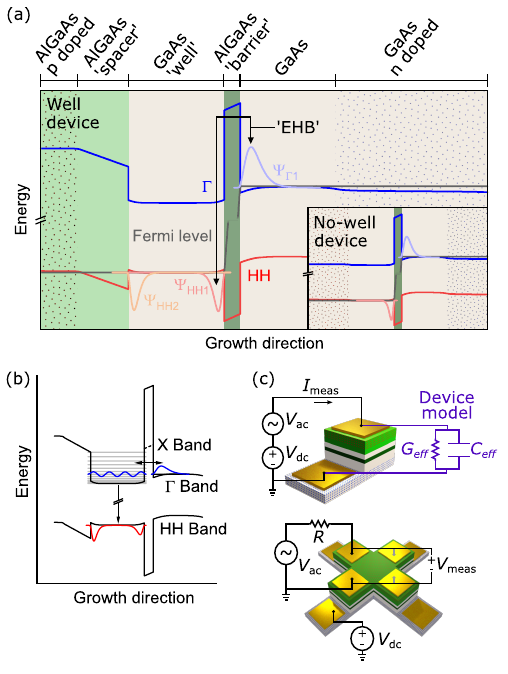}
	\caption{Layer structure and measurement setup. (a) Layers in `well' vs. `no-well' samples. AlGaAs layers are shown in green, while GaAs is in light grey. The barrier is an AlAs/GaAs superlattice. The band edges and wave functions are simulation results from nextnano \cite{Nextnano2007}; the bias is set to 1.65 V, and the well and p-doped layers are at the same potential. (b) Illustration of the resonant tunneling model. Electrons tunnel from the $\Gamma$ band on the n side (wave function shown in blue on the right) to discrete $\Gamma$ band states in the p-side well (blue wave function on the left). They then recombine with holes in the heavy-hole band on the p side of the device (red wave function on the left). Reproduced from \cite{Parolo2022}. (c) Upper: A voltage bias is applied between the contacts shown in yellow, with a small AC excitation voltage added to it. The resulting AC current is measured to extract the effective capacitance and conductance of the device. Lower: Cross-shaped samples with four contacts to each layer enable lateral resistivity measurements. Current is sent between two contacts, and the voltage across these contacts (2-point measurement) or across the remaining contacts (4-point measurement) is measured.}
	\label{fig:DeviceAndSetup}
\end{figure}

In this work, we study electronic transport in p-i-n GaAs/AlGaAs diodes with an additional AlGaAs barrier in the middle of the intrinsic layer. We distinguish between two classes of structures. In one, the p-doped layer is made of AlGaAs (`well samples'); in the other, it is made of GaAs (`no-well samples') (see Fig.~\ref{fig:DeviceAndSetup}a). `Well' here refers to the potential structure formed by the intrinsic GaAs layer situated between the AlGaAs barrier and AlGaAs doped layer. Note that in both cases the n-doped layer is GaAs; since the composition of this layer results in more minor changes to device behavior, a discussion of this property is relegated to the Supplement \cite{Sup}. In the no-well samples, capacitance measurements clearly indicate the formation of an EHB upon application of a bias corresponding to the GaAs band gap. This EHB consists of a 2D accumulation layer of holes on one side of the AlGaAs barrier and a 2D accumulation layer of electrons on the other side. Only upon lowering the temperature into the millikelvin regime do intriguing anomalies appear in interlayer transport measurements \cite{Davis2023}. In well samples, the behavior is more complex, as communicated in \cite{Parolo2022}, with interesting features present even at liquid helium temperatures. Here we focus on the well samples and for the first time present a detailed understanding of this complex behavior and what it means for the formation of EHBs.

In devices with a 10 nm barrier, prominent oscillations in both the capacitance and differential conductance are observed. We understand the conductance oscillations to be a result of resonant tunneling into quantum well states \cite{Parolo2022} (illustration provided in Fig.~\ref{fig:DeviceAndSetup}b). Here we extend the investigation of these devices, and show that the oscillations are strongly dependent on both temperature and the composition of an AlGaAs `spacer' grown between the p-doped layer and the intrinsic GaAs layer. In addition, entirely new transport measurements are performed, namely probes of lateral resistivity. These suggest that the p-doped layer and the holes in the well are not at the same potential, supporting the idea that there is a significant spacer resistance responsible for the observed capacitance oscillations. With this new understanding, we revisit and elucidate the hysteretic behavior observed by \cite{ParoloThesis} in samples with a thicker (15 nm) barrier. We then propose new heterostructure designs with modified spacers to facilitate charge transfer between the wells and the contacts, while also bringing the 2D charge layers closer to the barrier.


\section{Materials and Methods}

The semiconductor heterostructure devices presented here are grown by molecular-beam epitaxy. Rectangular or cross-shaped devices (see Fig.~\ref{fig:DeviceAndSetup}c) are then created by wet etching, after which ohmic contacts are deposited via electron-beam evaporation on the p-doped and n-doped layers. They are measured in a flow cryostat and a dipstick setup (1.3 K to room temperature possible). A voltage bias between an n contact and a p contact is applied, and the interlayer AC current response (capacitance and differential conductance) and intralayer resistance are measured as a function of this bias.

The capacitance and conductance are measured by adding a small AC voltage excitation to the DC voltage bias and measuring the AC current response using an IV converter and a lock-in amplifier. This requires that an amplitude and a frequency are selected for the excitation: 2.4 mV and 283 Hz were used for most measurements and unless otherwise specified. Using these values and the measurement results, the effective capacitance and conductance can be calculated. These effective values are the capacitance and conductance required to produce our measurement result if they were attached in parallel, as shown in Fig.~\ref{fig:DeviceAndSetup}c. A phase error of \textpm 1\textdegree\ is listed as the maximum of the lock-in. While this does affect the amplitude of the capacitance oscillations towards higher bias (where the in-phase component is larger) and for smaller frequencies (where the out-of-phase component is smaller), it does not alter their essential character for the frequency we usually use.

The intralayer resistance is measured by applying an AC voltage over two contacts to the same layer in series with a current-defining resistor. The voltage across these contacts (2-point) or the remaining contacts (4-point) is then measured with a lock-in amplifier. Note that, unlike in \cite{eisenstein1991,marty2023}, we contact the doped layers and not the 2DEG/2DHG directly. For this we use a frequency of 13.8 Hz, a voltage of 0.1 V, and a current-defining resistor of 1 M$\Omega$, unless otherwise noted. In the absence of any 2D charge layers, and assuming negligible charge transfer through the barrier, this technique simply measures the resistance of the doped layers. If, for example, a 2DEG exists near the barrier, the situation is more complicated. Assuming charge is transferred with insignificant resistance between the n-doped layer and the 2DEG, this technique now measures the resistance of the 2DEG in parallel with the n-doped layer. If this assumption does not hold, the measurement is influenced by the resistance from the 2DEG to the n-doped layer. Therefore, while we refer to this as the `intralayer resistance' measurement, in the devices we study it is often more complex.

Note that in this paper we restrict our discussion to those devices which show an initial lower-bias capacitance step followed by increases in the capacitance as a significant interlayer current begins to flow. For a brief discussion of the devices not included, see the Supplement \cite{Sup}.


\section{Framework}

\begin{figure}[!b]
	\centering
	\includegraphics[width=0.5\columnwidth]{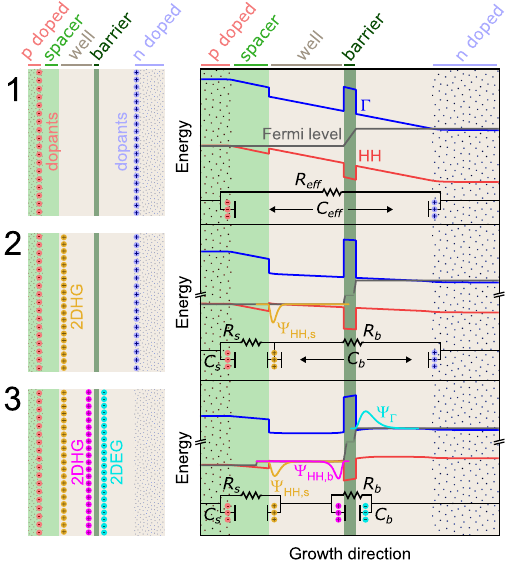}
	\caption{Illustration of the three different charge-distribution regimes used as a framework. The illustrations on the left show the charge layers present in each case. On the right, example band-edge diagrams are provided along with circuit models. A capacitor in parallel with a resistor is used to model the spacer ($R_s$, $C_s$) and the barrier ($R_b$, $C_b$) (the notation $G_s=R_s^{-1}$ and $G_b=R_b^{-1}$ is also used).}
	\label{fig:RegimeIllustration}
\end{figure}

In discussing these devices, it is useful to first create a framework by considering the different possible charge distribution situations, as well as the resulting DC currents in each. Firstly, at the lowest biases, we expect charges only at the edges of the doped layers. These would be ionized donors/acceptors in the n/p-doped layer, respectively, without free carriers to neutralize their charge, leading to small space-charge regions. As the bias is increased, we secondly expect that at a certain point holes can move across the spacer into the well. Assuming this is possible before the band edges become flat in the vicinity of the barrier, these holes will accumulate next to the spacer (with negative space charge still present in the p-doped layer). Thirdly, upon further increasing the bias, the band edges are expected to flatten out around the barrier, and shortly after this both holes and electrons can begin to accumulate next to the barrier---the `onset' of the EHB (with positive space charge no longer present in the n-doped layer). In the second and third situations, the potential difference across the spacer and that across the barrier must result in the same steady-state current. In the extreme case in which the barrier has a much higher resistance than the spacer, the p-doped layer and the well will be at the same potential. Note that the spacer could still have a high resistance, meaning long time scales for the well to reach the same potential as the p-doped layer. On the other hand, if significant current can flow across the barrier already at the EHB onset, the potential difference across the spacer must increase to match this current before our third situation can be achieved. The three different regimes outlined in this paragraph are illustrated in Fig.~\ref{fig:RegimeIllustration}, where the different charge layers present in each case are emphasized. Note that this is not an empirical framework, but rather an outline of the charge distribution situations we expect in these devices, based on fundamental electrostatics and semiconductor physics.

We now consider the AC current response of the devices. With the charges only in the doped layers, a simple parallel-plate capacitor is formed, and a corresponding purely out-of-phase current is measured. With holes accumulated in the well next to the spacer, we have a more complex situation that can be described by a parallel spacer capacitance and resistance in series with a parallel barrier capacitance and resistance (see Fig.~\ref{fig:RegimeIllustration}). Assuming $R_b$ is negligibly large, in the limit of small or large $R_s$, a capacitor of value $C_{b}$ or $(C_{s}^{-1}+C_{b}^{-1})^{-1}$, respectively, is realized. Intermediately, $R_{s}$ is significant and results in an in-phase component of the current. The final regime, in which the EHB has formed, requires consideration of the full model shown in Fig.~\ref{fig:RegimeIllustration}. Note that this model does not account for quantum capacitance, which could be included in future studies for improved accuracy at the cost of increased complexity. Mathematically, the circuit model is described by:

\begin{equation}
	\begin{split}
 		Y_{tot} &= G_{eff} + i \omega C_{eff}\\
 		G_+ &= G_s + G_b\\
 		C_+ &= C_s + C_b\\
 		G_{eff} &= \frac{[G_s G_b G_+] + [\omega^2(G_s C_b^2 + G_b C_s^2)]}{G_+^2 + (\omega C_+)^2}\\
 		C_{eff} &= \frac{[C_s G_b^2 + C_b G_s^2] + [\omega^2(C_b C_s C_+)]}{G_+^2 + (\omega C_+)^2}
	\end{split} \label{eq:1}
\end{equation}
\smallskip

Here $i$ is the imaginary unit, and $G_{eff}$ and $C_{eff}$ are the effective conductance and capacitance values, respectively. Both effective values become constant in the limit of very small or large frequencies. In the intermediate regime, they will change with frequency, excluding certain special cases. One such case is if $G_s \rightarrow \infty$, which is ideal from the perspective of studying EHBs (no resistance to the contacts). Then $G_{eff}=G_b$ and $C_{eff}=C_b$ always applies. The no-well devices come closer to this condition \cite{Davis2023}. Allowing $G_s$ to take finite values, but keeping $G_s >> G_b$, we can consider what effect this imperfection has on the AC current response. We write $C_{eff} = C_b + \Delta C$, where $\Delta C$ is the effective capacitance's deviation from $C_b$, and consider the low-frequency limit. The resulting relation is $\Delta C \approx C_b (-2 G_b/G_s)$. The deviation thus tracks the negative of $G_b$ and is enhanced by smaller values of $G_s$. This can be intuitively understood by considering two resistances in series, one constant (spacer) and one oscillating with the voltage applied across it (barrier). In this case, the way the total applied voltage is divided between the two resistors changes as it is increased. If the variable resistance is decreasing, it will take a smaller portion of the applied voltage. A capacitor in parallel with this variable resistor will correspondingly draw less charge. Thus for an oscillating differential conductance (with applied voltage) we expect capacitance oscillations matching the negative of this conductance.

\begin{figure}[!t]
	\centering
	\includegraphics[width=0.5\columnwidth]{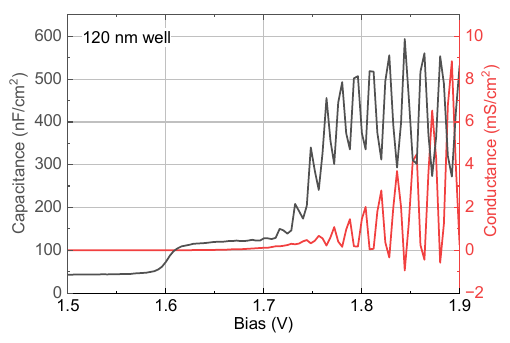}
	\caption{Capacitance and interlayer conductance of a sample with a 120 nm well and a 10 nm barrier. The Al content of the spacer is 30\% and the temperature is around 5 K.}
	\label{fig:ExampleCapacitance}
\end{figure}

Before moving on to the main results, we first present one example capacitance vs. DC bias measurement and relate it to the framework developed above. This is shown in Fig.~\ref{fig:ExampleCapacitance}. For biases under 1.6 V, we measure a relatively constant capacitance, which agrees well with the geometric capacitance of $\text{41 nF/cm}^2$ calculated for charge layers at the edges of the doped layers (our first regime). The capacitance step around 1.6 V reflects charges moving closer to the barrier. The value of the subsequent plateau reflects $C_s$ in series with $C_b$ after the formation of the EHB, suggesting the device is in the third regime, with $R_s$ and $R_b$ still very large. The intermediate second regime is likely not visible in the capacitance due to the high value of $R_s$. As the bias is increased to values above 1.7 V, $R_s$ appears to drop. The capacitance can then rise towards $C_b$, but is modulated by $G_b$. This is in accordance with equation given above, with the capacitance dropping as the conductance peaks, and vice versa.

\section{Results and Discussion}

\subsection{Capacitance Oscillations}

We first present the results of measurements of samples with a 10 nm AlGaAs barrier. The temperature dependence was measured on a set of five samples. These samples had spacers with a width of 30 nm and an Al content of 30\%. They had well widths from 45 to 150 nm. In contrast, when studying the effect of the spacer properties, the well width was fixed at 45 nm. The spacer width was also fixed, at 30 nm, and its Al content was varied between 10\% and 30\% for three different samples.

\subsubsection{Interlayer Transport}

\begin{figure}[!t]
	\centering
	\includegraphics[width=0.5\columnwidth]{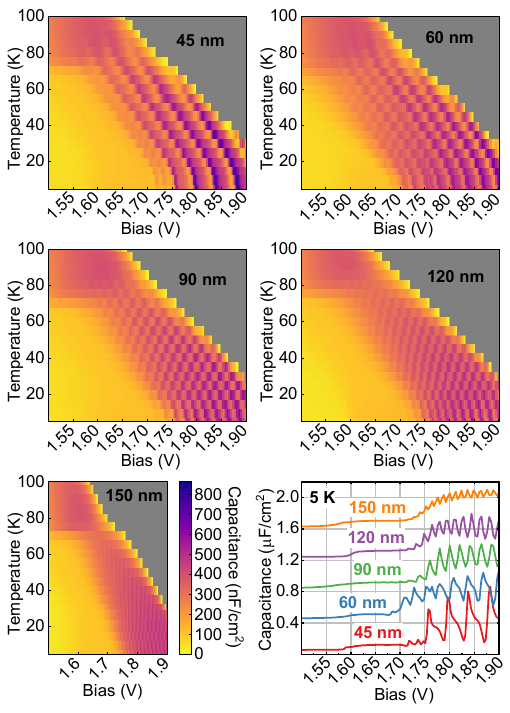}
	\caption{Capacitance of all five 30\% Al samples vs. temperature and bias voltage. The width of the GaAs layer on the p side is indicated on the top right of each plot. Only the positive capacitance values are shown, with negative values colored gray, and nearest-neighbor interpolation is used. The shift of individual peaks with increasing temperature is clearly visible in the 45 nm and 60 nm plots. For larger well widths, a higher resolution in temperature would be needed to observe this trend. Cross-sections at 5 K are shown in the bottom right plot, where the differing oscillation periods are evident. The curves have been offset for clarity.}
	\label{fig:TemperatureDependence}
\end{figure}

Conductance oscillations are observed in all the samples with a 10 nm barrier and a p-side well. The period of the oscillations is larger for thinner wells, and it is consistent with the energy-level spacing of $\Gamma$-band states in the well. We thus propose that the oscillations are a result of the resonant tunneling of electrons from the n-side of the device to the p-side well, as described in \cite{Parolo2022} and shown in Fig.~\ref{fig:DeviceAndSetup}b. These conductance oscillations are accompanied by oscillations in the capacitance, and both are gradually suppressed in all samples as temperature is increased, as shown in Fig.~\ref{fig:TemperatureDependence}. One might expect that for thinner wells, with consequently larger energy-level spacing, the oscillations persist to higher temperatures. This is, however, not what is observed, with the oscillations disappearing around 90 K for all samples. Additionally, the oscillations in both the capacitance and conductance tend to be enhanced in amplitude towards higher bias voltages. As the bias is further increased, there comes a point, which becomes progressively lower with increasing temperature, where the conductance increases rapidly while the capacitance curves downwards and eventually becomes negative (discussed later).

Before the oscillations appear, a small step up in the capacitance to around $\text{100 nF/cm}^2$ is observed in all samples (orange regions in Fig.~\ref{fig:TemperatureDependence}). We interpret this as the onset of the EHB, but with a large spacer resistance resulting in an AC current response reflecting $C_s$ in series with $C_b$ (as also mentioned above). As $R_s$ and $R_b$ become lower with bias the oscillations begin. As we can expect all resistances to decrease with increased temperature, the observed leftwards shift of the capacitance features with increasing temperature makes sense. Note that the expected decrease of the band-gap energy over this temperature range is too small to account for this shift. The accompanying suppression of the capacitance oscillations follows from the suppression of those in the conductance. In the resonant tunneling model, this should occur when significant tunneling can take place to multiple energy levels at once. Comparing $k_B T$, with $k_B$ the Boltzmann constant and $T$ the temperature, to the quantum well energy level spacing does not predict the $\approx$ 90 K limit. For example, for a well width of 45 nm, with around 41 meV between energy levels, $\frac{3}{2} k_B T$ only exceeds the energy-level difference around 316 K. The reason for this discrepancy remains an open question but could be a result of lifetime broadening of the well levels as temperature is increased.

\begin{figure}[!t]
	\centering
	\includegraphics[width=0.5\columnwidth]{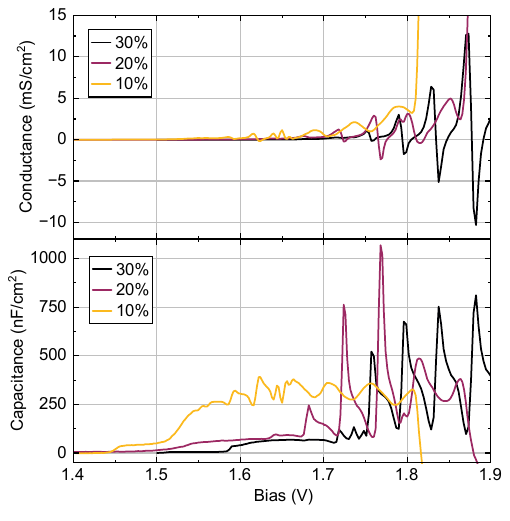}
	\caption{Differential conductance and capacitance as a function of applied bias for three samples with different percentages of Al in the spacer (4.2 K, 2.5 mV AC excitation). The capacitance curves are offset to align the low-bias values.}
	\label{fig:SpacerMeas}
\end{figure}

To probe the effect of the spacer resistance more directly, three different samples were grown with 10\%, 20\%, and 30\% Al in the spacer. A lower potential barrier should form between the p-doped layer and the well with lower Al content, resulting in a lower $R_s$ value, which is indeed what we find. As with increasing temperature, the features in the capacitance are shifted to the left with decreasing Al content, as seen in Fig.~\ref{fig:SpacerMeas}. In addition, for the 10\% sample, the conductance oscillations are suppressed. Here too decreased confinement and resulting lifetime broadening could be responsible. Another noteworthy fact about the 10\% sample is that the onset is shifted so much as to be lower than the band-gap energy (around 1.519 V). This implies that the spacer resistance is low enough to observe a capacitance rise before the EHB has formed, reflecting rather the accumulation of holes in the well adjacent to the spacer (the second situation in our framework). The formation of the EHB can then be seen clearly around 0.07 V later, with a capacitance step that reflects $C_b$.

\subsubsection{Intralayer Transport}

\begin{figure}[!b]
	\centering
	\includegraphics[width=0.5\columnwidth]{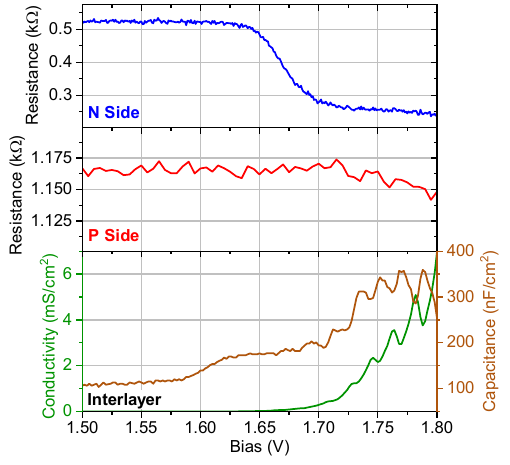}
	\caption{Intralayer resistance (2-point) on both the n and p sides of a 90 nm well device. The spacer Al content is 30\% and width is 30 nm. These are compared to interlayer measurements on the same device (141 Hz, 2.5 mV AC excitation). All at 4.2 K.}
	\label{fig:LateralMeas}
\end{figure}

As described above, intralayer resistance measurements were performed on cross-shaped samples. Assuming negligible spacer resistance and infinite barrier resistance, these measurements extract the resistance of the doped layer in parallel with the 2D layers (the 2DEG or the 2DHGs). We thus expect a significant drop in resistance at the bias where the high-mobility 2D systems begin to form. This is what is seen in no-well samples, with pronounced drops in the intralayer resistivity occurring on both the n and p sides around the band-gap voltage \cite{Davis2023}. However, if we instead have a substantial spacer resistance, this drop can be delayed, since the current can no longer be easily redirected through the higher-conductivity 2D channel(s) (until $R_s$ becomes, with increased bias, small enough). We indeed see very little change in the p side resistivity with bias, while observing a large drop on the n side, as shown in Fig.~\ref{fig:LateralMeas}. This implies a spacer resistance significantly larger than that of the p-doped channel, which is on the order of 1 k$\Omega$. These measurements allow us to obtain far more direct evidence of the spacer resistance.

\subsubsection{Frequency Dependence}

Up until this point, our discussion has been qualitative. We now attempt a more quantitative treatment. As discussed above, depending on the spacer and barrier properties, a frequency dependence is predicted. A frequency sweep was thus performed from 125 to 1000 Hz on a sample with a 45 nm well and a 30 nm spacer with 30\% Al. The behavior of this device was then compared with a model generated from equation \eqref{eq:1} and estimates of the circuit elements $R_s$ (or $G_s$), $R_b$ (or $G_b$), $C_s$, and $C_b$.

\begin{figure}[!b]
	\centering
	\includegraphics[width=0.5\columnwidth]{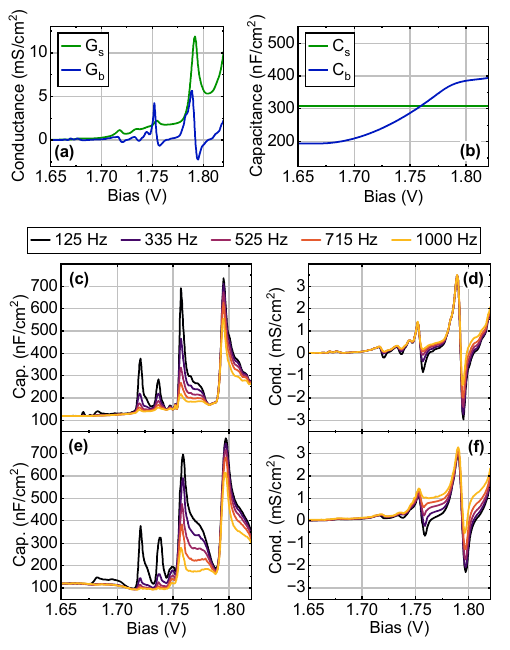}
	\caption{Modeling frequency behavior in a 45 nm well sample with a 30\% Al, 30 nm spacer. (a) Estimates for $G_s$ and $G_b$. (b) Estimates for $C_s$ and $C_b$. (c) Modeled capacitance. (d) Modeled conductance. (e) Capacitance data. (f) Conductance data. Both at 4.2 K, with a 2.5 mV AC excitation.}
	\label{fig:FrequencyAnalysis}
\end{figure}

Fig.~\ref{fig:FrequencyAnalysis}a shows the estimates we use for $G_s$ and $G_b$, while Fig.~\ref{fig:FrequencyAnalysis}b shows the estimates we use for $C_s$ and $C_b$. These values are chosen to be realistic while still providing a good fit to the data. For details on the estimation of these parameters, see the Supplement \cite{Sup}. With the values for all four circuit elements present in our model, we can simply use equation \eqref{eq:1} to calculate the effective capacitance and conductance vs. bias. The result of plugging these values into equation \eqref{eq:1} is shown in Fig.~\ref{fig:FrequencyAnalysis}c-d, with the data in Fig.~\ref{fig:FrequencyAnalysis}e-f. Although the match is not exact, the main features of the data are reproduced by the model.

While the mechanism outlined above can explain enhancement and reduction of the effective capacitance, it cannot explain the negative effective capacitance values observed. Negative capacitance is observed in diodes at high forward currents \cite{Barna1971}. This is thought to be a result of `conductivity modulation': the reduction of the bulk resistance as the diode current is increased. The essential point here is that there exists a series resistance that decreases with increased current \cite{Bisquert2011}. Our estimated series resistance does indeed behave this way and so may be responsible for the negative capacitance values observed. However, understanding this would require more careful study and measurement with very low phase errors, so here we offer only a tentative suggestion.


\subsection{Hysteresis}

We now turn to p-i-n devices with an increased barrier thickness of 15 nm. The spacer width is again 30 nm, with an Al content of 30\%. Despite the modest change in the layer structure, these samples exhibit markedly different behavior compared to those with a thinner barrier. We no longer observe the capacitance and conductance oscillations, but rather see novel hysteretic effects.

\begin{figure}[!t]
	\centering
	\includegraphics[width=0.5\columnwidth]{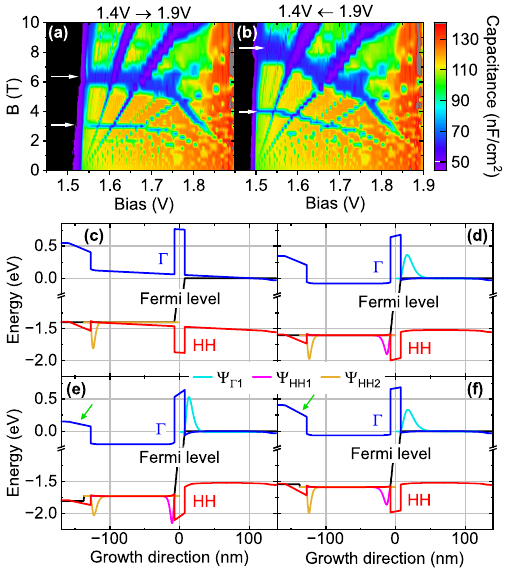}
	\caption{Hysteresis in a 15 nm barrier sample with a 30 nm, 30\% Al spacer. (a) The capacitance with the bias swept up. The AC excitation amplitude is 1 mV, and the frequency is 78 Hz. The temperature is 25 mK. (b) As before, with the bias swept down. (c) Band edges at 1.4 V. (d) At 1.6 V. (e) At 1.8 V, with the well at a lower potential. (f) At 1.54 V, with the well at a higher potential.}
	\label{fig:Hysteresis}
\end{figure}

In these samples, the capacitance step corresponding to the EHB onset occurs at different biases depending on in which direction the bias is swept. On the up sweep, the bias at the capacitance step is similar to the band-gap energy. On the down sweep, this bias is shifted down by around 0.05 V. Performing these sweeps over a range of magnetic field values reveals a rich pattern in the hysteresis, as was done in \cite{ParoloThesis} and is reproduced in Fig.~\ref{fig:Hysteresis}a \& b.

A Landau fan is observed in the 2D magnetic field and bias plots. The most obvious source for the fan is 2D charge layers accumulated on either side of the barrier (assuming that these charge layers have equal densities, they will produce only one fan, not two). In addition to this, there is another set of dips which does not match the typical fan pattern, but rather extends horizontally across the plots or moves downwards in magnetic field with increasing bias. As we explain in the following paragraphs, this is likely a signature of the 2D hole layer accumulated next to the spacer.

The interpretation described above is supported by self-consistent nextnano \cite{Nextnano2007} simulations and a consideration of the capacitance values observed. To begin, note that the capacitance remains slightly over 100 nF/cm$^2$ from the onset bias to around 1.85 V (here we consider the measurement at 0 T). This value is consistent with $C_s$ in series with $C_b$. Additionally, the capacitance step on the up sweep occurs near the band-gap energy. These facts suggest that both $R_s$ and $R_b$ are high, but $R_b$ far exceeds $R_s$, so the spacer experiences a very low potential bias and the onset is not delayed. This is in contrast to the samples with a 10 nm barrier, where $R_b$ is presumably much lower. Simulations in nextnano with the entire p side (i.e., well and p-doped layer) at the same Fermi energy predict a density of $3 \cdot 10^{11}\ \text{cm}^{-2}$ for the wave function next to the spacer, in good agreement with the density extracted from the horizontal lines in Fig.~\ref{fig:Hysteresis}a (assuming spin splitting is not resolved). The nextnano predictions are also consistent with the fan density, now considering the wave functions next to the barrier and assuming spin splitting is resolved. Some results of these self-consistent simulations are shown in Fig.~\ref{fig:Hysteresis}c-f, with Fig.~\ref{fig:Hysteresis}c showing what the band edges look like prior to onset, and Fig.~\ref{fig:Hysteresis}d showing the situation with the EHB formed and the well and p-doped layer at the same potential.

This explains the behavior on the up sweep until a bias around 1.7 V. After this point, the previously horizontal lines begin to curve downwards. We propose that this reflects the beginning of current flow through the barrier (or, in other words, the decrease of $R_b$). The spacer and barrier must permit the same current flow, so a potential drop forms from the p-doped layer to the well, reducing the potential barrier presented by the spacer and decreasing its resistance. The reduced band bending of the spacer requires a reduced accumulation of holes next to it. This is shown in Fig.~\ref{fig:Hysteresis}e (the spacer slope is highlighted with a green arrow).

Following this line of reasoning, the hysteresis emerges as a result of restricted charge transfer from the well out towards the p-doped layer. Decreasing the bias shows the once-horizontal lines moving upwards (Fig.~\ref{fig:Hysteresis}b), suggesting a buildup of holes against the spacer, transferred from the hole layer next to the barrier. Fig.~\ref{fig:Hysteresis}f shows this, with the slope of the band edges at the spacer now enhanced and the potential in the well higher (i.e., the electron energy lower) than that in the p-doped layer. Eventually this potential difference becomes high enough for the charges to move out of the well. Note that with a barrier thin enough to permit current flow these charges would be able to simply recombine with the electrons on the other side. Not only is this behavior interesting in its own right, but it also elucidates the role of the spacer and its interplay with the barrier.


\section{Outlook}

In this paper, we demonstrated how the interplay between the AlGaAs barrier and spacer in p-i-n diode EHB devices can result in a variety of phenomena---particularly, capacitance oscillations and hysteresis. These behaviors could be tuned by adjusting the properties of the AlGaAs layers. Particularly, the spacer resistance could be significantly lowered by reducing the Al content. For an AlGaAs spacer with 10\% Al, a capacitance step suggesting charge accumulation in the well was observed prior to the band-gap voltage. In the future, the spacer Al content could be further reduced or the effects of thinner spacers studied, with the aim of achieving an even lower spacer resistance, while still having a confining well.

\begin{figure}[!b]
	\centering
	\includegraphics[width=0.5\columnwidth]{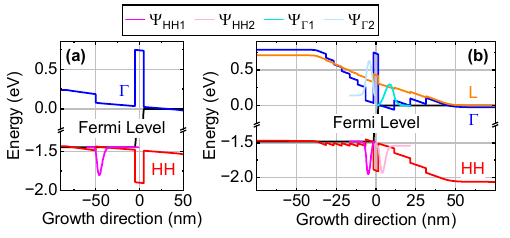}
	\caption{Band edge simulations of (a) the 10\% Al content device and (b) a proposed new device with more closely spaced electron and hole layers.}
	\label{fig:Outlook}
\end{figure}

In addition, the accumulation of charges prior to the band-gap voltage suggests another avenue to explore for the realization of closely spaced EHBs. Prior to the flat-band condition, the wave function in the well faces a heavy-hole band edge sloping downwards towards the barrier (see Fig.~\ref{fig:Outlook}a, where the simulated bands for the 10\% Al spacer device are shown). This could resist the flow of charge from the well across the barrier to the n side of the device---there are no states available to tunnel into on the other side of the barrier, with the first state on the n side being offset in energy. Thus if the EHB could be formed at this point, there is the potential that much thinner barriers could be used without facing the issue of large interlayer leakage currents. We present an example device design to achieve this (Fig.~\ref{fig:Outlook}b). Here the p-doped AlGaAs layer has a relatively high Al content, around 60\%. As the spacer is grown, the Al content is stepped down over 30 nm until it reaches around 20\% (alternatively, the Al content could be smoothly decreased as growth progresses). In this way, the potential barrier presented by the spacer is similar to that of the measured 10\% Al device, and charge accumulation theoretically occurs before the flat-band condition, but the well can be made much thinner. A similar strategy is then used on the n side of the device. Here a limit is placed on the Al content of the n-doped layer by the lowering in energy of the L band. The well on the p side is made thinner to further inhibit tunneling of the electrons, which are lighter than the heavy holes. Many properties are available here to further tune and study.

We conclude by relating these devices back to the no-well devices for context. The no-well devices are in a sense ideal for the study of EHBs, due to the much lower resistance from the p-doped layer to the 2DHG, and the ability to achieve closely spaced EHBs \cite{Davis2023}. However, the lack of wells means there are limited options available for further decreasing the EHB interlayer spacing. The well devices, on the other hand, present a rich playground of parameters for further study. With the measurements made and analytical techniques developed in this paper to understand this class of AlGaAs-spacer devices, the groundwork is laid to experiment with new devices with the aim of creating more strongly interacting EHBs. With stronger electron-hole interactions, these devices would have the potential to extend further into the excitonic regime.


\section{Acknowledgments}

We thank Peter Märki for sharing his technical expertise and providing valuable support. We acknowledge financial support from the Swiss National Science Foundation (SNSF) and the National Center of Competence in Science “QSIT—Quantum Science and Technology.”


%

\end{document}


\title{Supplemental Material: Charge Transfer Dynamics in an Electron-Hole Bilayer Device: Capacitance Oscillations and Hysteretic Behavior}

\author{M. L. Davis}
\affiliation{Solid State Physics Laboratory, ETH Z\"urich, CH-8093 Z\"urich, Switzerland}
\affiliation{Quantum Center, ETH Z\"urich, CH-8093 Z\"urich, Switzerland}
\author{S. Parolo}
\affiliation{Solid State Physics Laboratory, ETH Z\"urich, CH-8093 Z\"urich, Switzerland}
\author{S. Agostini}
\affiliation{Solid State Physics Laboratory, ETH Z\"urich, CH-8093 Z\"urich, Switzerland}
\author{C. Reichl}
\affiliation{Solid State Physics Laboratory, ETH Z\"urich, CH-8093 Z\"urich, Switzerland}
\affiliation{Quantum Center, ETH Z\"urich, CH-8093 Z\"urich, Switzerland}
\author{W. Dietsche}
\affiliation{Solid State Physics Laboratory, ETH Z\"urich, CH-8093 Z\"urich, Switzerland}
\affiliation{Max-Planck-Institut f\"ur Festk\"orperforschung, D-70569 Stuttgart, Germany}
\author{W. Wegscheider}
\affiliation{Solid State Physics Laboratory, ETH Z\"urich, CH-8093 Z\"urich, Switzerland}
\affiliation{Quantum Center, ETH Z\"urich, CH-8093 Z\"urich, Switzerland}

\maketitle

\section{N-Side Composition}

In the main text, we restrict ourselves to samples with only GaAs on the n-side of the device. Other devices were grown with n-doped AlGaAs, for which minor changes in the capacitance traces were observed. This can be seen in \cite{ParoloThesis}, Fig. 4.5. Here a device with the n-doped layer made of AlGaAs is compared to devices for which it is made of GaAs. The AlGaAs doped layer is separated from the intrinsic GaAs layer by a 30 nm spacer and the Al content is around 19\%. A small step is observed in the AlGaAs-doping device at low biases, which is attributed to charges moving into the well on the n side and settling at the spacer edge. We additionally highlight that this is consistent with nextnano simulations, where occupation of a wave function next to the spacer is seen for a bias of 0.8 V.

This implies that the n-side spacer resistance is largely insignificant for the spacer width and composition in the device just discussed. Another device with a much wider n-side spacer (70 nm) was also grown, for which this may no longer be the case. The Al content for this device is also around 19\%. This sample lacks a pronounced drop in the n-side intralayer resistivity, implying the spacer resistance is now substantial (under the assumption that the 2DEG mobility is not drastically reduced). This device is discussed further in its own section (`Current Jumps'), since it exhibits interesting, unusual behavior.

We can conclude that the device behavior is much less sensitive to the properties of the n-side spacer (compared to the p-side spacer) and that the electrons can easily move across it for the smaller widths (30 nm) actually required for the spacer.

\section{Additional Devices}

\begin{figure}[!ht]
	\centering
	\includegraphics[width=0.5\columnwidth]{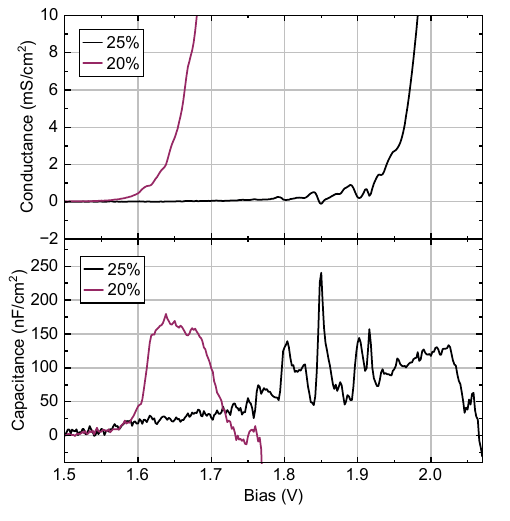}
	\caption{Capacitance and differential conductance of additional samples with variations in the spacer composition (4.2 K, 2.5 mV AC excitation, capacitance curves offset). The spacer width is 30 nm and the barrier width is 10 nm.}
	\label{fig:OtherSpacerTests}
\end{figure}

Here we present additional devices that had behavior deviating from that of the majority of the devices tested. The first of these, grown as part of the series studying the effect of spacer Al percentage, are shown in Fig.~\ref{fig:OtherSpacerTests}. One has an Al content of 20\%, just as the one presented in the main text. However, oscillations are not observed, and the interlayer current shoots up at a relatively low bias, and in unison with the first increase in the capacitance. Here we have a plausible explanation: for this sample (compared to the sample in the main text), the edge of the n-doped layer was grown at a higher temperature. This could have allowed donors to migrate towards the barrier, causing the interlayer leakage current as the bias is increased. Decreasing the temperature to around 500\textdegree C seems to have resolved the issue. We also grew a sample with an Al content of 25\%. This sample exhibits recognizable features, with an initial lower-bias capacitance step followed by oscillations at higher bias. However, both the capacitance and conductance signals seem to be suppressed. The reason for this is unclear.

We briefly mention one more device, which has been presented in \cite{ParoloThesis}, Fig. 4.6b. This device, despite the p-doped layer being made of AlGaAs and the presence of an accompanying spacer, showed only minor deviations from a similar device with the p-doped layer made of GaAs. This would imply low resistance of the spacer in this structure. At this point, we do not have an explanation for this anomaly. 

\section{Current Jumps}

\begin{figure}[!ht]
	\centering
	\includegraphics[width=0.5\columnwidth]{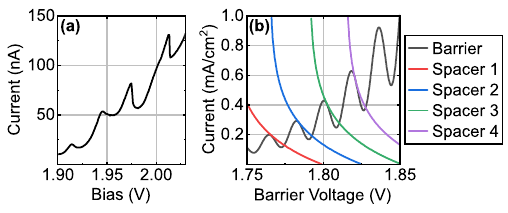}
	\caption{(a) Current jumps in a sample with asymmetric wells (4.2 K). (b) Illustration of how current jumps might occur.}
	\label{fig:CurrentJumps}
\end{figure}

In this section, a device with doping done in AlGaAs on both the n and p sides is briefly discussed. While it does not belong to the class of devices discussed in the main text, it provides another example of the interesting behavior that can be found in the broader class of p-i-n diode devices---namely, current jumping. No spacer is present on the p side (doped layer has 29\% Al), and a 70 nm spacer is present on the n side (doped layer has 19\% Al). In this sample, we observed a peculiar frequency dependence at certain biases. Upon inspecting the behavior of the current in response to a square wave with an oscilloscope, jumps were observed. This can also be seen by measuring the current in a bias sweep, as shown in Fig.~\ref{fig:CurrentJumps}a. The peaks below 1.95 V are relatively smooth, while above the current drops suddenly after peaking. This may again be a consequence of the relationship between the barrier and the spacer resistance. An example model that could result in current jumping is given in Fig.~\ref{fig:CurrentJumps}b. Four different example spacer curves are shown for different total biases. The steady-state situation occurs when the spacer and barrier curves intersect. For the examples shown, this occurs at two distinct points. Thus, with a good understanding of the spacer and barrier current-voltage characteristics, devices with very high negative differential conductance or even current jumps can be engineered.

\section{Frequency Dependence: Parameter Estimation}

In this section, we discuss the estimation of $R_s$ (or $G_s$), $R_b$ (or $G_b$), $C_s$, and $C_b$ for the analysis of the frequency dependence that is presented in the main text.

\begin{figure}[!ht]
	\centering
	\includegraphics[width=0.5\columnwidth]{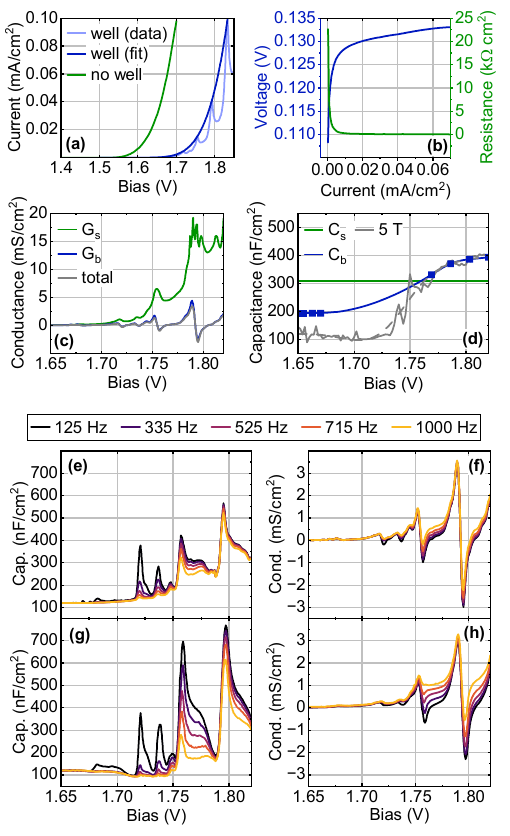}
	\caption{Modeling frequency behavior in a 45 nm well sample with a 30\% Al, 30 nm spacer. This figure shows alternative parameter estimations. (a) Current in the no-well and well samples used to construct (b) the current-voltage relationship for the spacer. (c) Estimates for $G_s$ and $G_b$. The total conductance is used to calculate $G_b$ from $G_s$. (d) Estimates for $C_s$ and $C_b$ (same as in main text). The 5 T data is shown in grey, with the smoothed data displayed as a dashed line. The blue squares show the points used to construct the spline. (e) Modeled capacitance. (f) Modeled conductance. (g) Capacitance data (same as in main text). (h) Conductance data (same as in main text). Both at 4.2 K, with a 2.5 mV AC excitation.}
	\label{fig:FrequencyAnalysisAlternative}
\end{figure}

Firstly, an estimate of the spacer resistance is obtained by comparing the current vs. bias of the 45 nm well sample to the current in a sample without a well (p-doped layer in GaAs). A polynomial fit is done to the peaks in the well current and taken as an estimate of the current vs. voltage that would occur in a no-well sample with the spacer resistance added (Fig.~\ref{fig:FrequencyAnalysisAlternative}a). The current/voltage characteristic of the spacer can then be extracted by taking the bias difference of the no-well curve and the fit curve at each current. This curve is differentiated to obtain the spacer resistance as a function of current (Fig.~\ref{fig:FrequencyAnalysisAlternative}b), which can easily be used to get $G_s$ at each bias. Differentiating the original well current vs. bias we get the DC total conductance and finally additionally extract $G_b$ at each bias (using that the total conductance is $G_s$ in series with $G_b$). Note that it is the oscillations in the total conductance that cause corresponding oscillations in the estimated values of $G_s$ and $G_b$. This is presented in Fig.~\ref{fig:FrequencyAnalysisAlternative}c. Note that smoothing is done in these steps in an attempt to prevent features appearing simply as a result of e.g. noise enhanced by differentiation. This is the estimate used in the model shown in Fig.~\ref{fig:FrequencyAnalysisAlternative}e-f. At biases above around 1.75 V, the modeled capacitance has a reduced spread with frequency compared to the data (Fig.~\ref{fig:FrequencyAnalysisAlternative}g-h). This suggests that the rough estimation method presented here results in an overestimate of $G_s$. For the model shown in the main text, $G_s$ is therefore adjusted to produce a better fit to the data.

In order to estimate the capacitances $C_s$ and $C_b$, the effective capacitance is measured in an in-plane magnetic field of 5 T. In such high magnetic fields, the conductance and capacitance oscillations are suppressed, as first observed in \cite{Parolo2022}. This is likely due to the reduction of resonant tunneling across the barrier by the parallel field. With $R_s$ low enough, this measurement yields a good estimate of $C_b$. This is a decent assumption only for higher biases (above around 1.75 V in Fig.~\ref{fig:FrequencyAnalysisAlternative}d). We additionally estimate a constant spacer capacitance, taken as the parallel-plate capacitance value corresponding to a distance of 30+7 nm. 30 nm is the spacer width and 7 nm is an estimate for the hole wave function offset from the spacer edge (based on nextnano \cite{Nextnano2007} simulations). The low-bias $C_b$ is then taken as the value that, taken with $C_s$, produces the measured total series capacitance at the onset (around 1.65 V in Fig.~\ref{fig:FrequencyAnalysisAlternative}d). A spline is fit to interpolate between these two regimes. This estimate is used in Fig.~\ref{fig:FrequencyAnalysisAlternative} and in the main text.